\documentclass[prl,nofootinbib,nobibnotes,twocolumn,superscriptaddress]{revtex4-1}

\usepackage{graphicx}
\usepackage{amsmath,amsfonts,amsthm,amssymb} 
\usepackage{float}
\usepackage{bbold}
\usepackage{mathtools}
\usepackage{gensymb}
\usepackage{times}

\begin{document}

\title{Topological transitions induced by cavity-mediated interactions in photonic valley-Hall metasurfaces}

\author{Charlie-Ray Mann}
\email{cm433@exeter.ac.uk}
\affiliation{School of Physics and Astronomy, University of Exeter, Exeter, EX4 4QL, United Kingdom.}

\author{Eros Mariani}
\affiliation{School of Physics and Astronomy, University of Exeter, Exeter, EX4 4QL, United Kingdom.}

%==================================================
%------------------------------------ABSTRACT---------------------------------%
%==================================================

\begin{abstract}
\noindent Topological phases of light exhibit unique properties beyond the realm of conventional photonics. In particular the valley-Hall topological insulator has been realized in a variety of photonic structures because it can be easily induced by breaking certain lattice symmetries. However, the valley-Chern numbers are usually fixed by design and the corresponding topological edge states are forced to propagate in a fixed direction. Here, we propose a mechanism to induce topological transitions via accidental Dirac points in metasurfaces composed of interacting dipole emitters/antennas. For a fixed arrangement of dipoles, we show that the topological phase depends critically on the surrounding electromagnetic environment which mediates the dipole-dipole interactions. To access different topological phases we embed the metasurface inside a cavity waveguide where one can tune the dominant dipolar coupling from short-range Coulomb interactions to long-range photon-mediated interactions by reducing the cavity width; this results in a topological transition characterized by an inversion of the valley-Chern numbers. Consequently, we show that one can switch the chirality of the topological edge states by modifying only the electromagnetic environment in which the dipoles are embedded. This mechanism could have important implications for other topological phases such as photonic higher-order topological insulators.
\end{abstract}

\maketitle

%==================================================
%---------------------------------MAIN TEXT------------------------------------%
%==================================================

\noindent \textbf{Introduction.---} Photonic topological insulators have revolutionized our ability to manipulate light in unprecedented ways, such us enabling the robust transport of light through certain defects and disorders without backscattering \cite{Lu2014a,Khanikaev2017,Ozawa2019,Wang2009,Hafezi2011,Fang2012,Khanikaev2013,Rechtsman2013,Hafezi2013,Wu2015,Cheng2016,Ma2016}. Within this paradigm of topological photonics, the valley-Hall topological insulator (VHTI) has attracted considerable interest because it does not require external fields \cite{Wang2009} or temporal modulation \cite{Fang2012} to break time-reversal symmetry; instead, the VHTI phase is easily induced by breaking certain lattice symmetries \cite{Ma2016}. Due to this simple geometrical origin, VHTIs have been successfully realized in a variety of photonic structures at microwave \cite{Wu2017,Gao2017,Gao2018}, terahertz \cite{Zeng2020,Yihao2020}, and optical \cite{Noh2018,Shalaev2019,Xin-Tao2019,Mehrabad2019,Barik2019} frequencies. These works begin with hexagonal lattices that exhibit deterministic Dirac points \cite{Lu2014} and then engineer a topological gap by breaking inversion and/or mirror symmetries. These symmetry-breaking perturbations generate localized Berry curvature near the inequivalent valleys and the corresponding VHTI phase is characterized by non-trivial valley-Chern numbers. An interface separating two regions with opposite valley-Chern numbers supports valley-locked chiral edge states which can be exploited to transport light through sharp bends and certain disorders which do not mix the valleys --- a tantalizing prospect that could have crucial implications for photonic devices such as topological waveguides \cite{Wu2017,Gao2017,Gao2018,Zeng2020,Yihao2020,Noh2018,Shalaev2019,Xin-Tao2019,Mehrabad2019,Barik2019}, directional antennas \cite{Gao2017}, topological splitters \cite{Wu2017}, chiral quantum optical interfaces \cite{Mehrabad2019,Barik2019}, and topological lasers \cite{Zeng2020}.

The main drawback with photonic VHTIs is that the topological phase is usually dictated by the symmetry-breaking perturbation that is imprinted into the lattice design. To deterministically change the valley-Chern numbers one needs to modify the symmetry-breaking parameter to induce a topological transition, passing through a critical point where the symmetry that gives rise to the deterministic Dirac points is restored. However, it is usually difficult, if not impossible, to reconfigure the lattice parameters after the photonic structure has been fabricated.  Consequently, the valley-Chern numbers are generally fixed by design and therefore the topological edge states are forced to propagate in a fixed direction \cite{Wu2017,Gao2017,Gao2018,Zeng2020,Yihao2020,Noh2018,Shalaev2019,Xin-Tao2019,Mehrabad2019,Barik2019}. This raises an important question: how can we induce a topological transition and thus switch the chirality of the topological edge states \emph{without modifying the symmetry-breaking parameter}?

Here we propose an alternative mechanism to induce topological transitions via \emph{accidental Dirac points} in kagome metasurfaces composed of subwavelength arrays of dipole emitters/antennas. The collective dynamics of the interacting dipoles results in subradiant polariton states that are bound to the metasurface \cite{Mann2018,Mann2020}. For a metasurface with a fixed symmetry-breaking perturbation, we show that the corresponding valley-Chern numbers depend critically on the nature of the dipole-dipole interactions which are mediated by the surrounding electromagnetic environment. To access different topological polariton phases, we embed the metasurface inside a cavity waveguide where one can tune the dominant dipolar coupling from short-range Coulomb interactions to long-range photon-mediated interactions by reducing the cavity width \cite{Mann2018,Mann2020}. Crucially, at a critical cavity width, the photon-mediated interactions eliminate the topological band gap induced by the Coulomb interactions; this results in accidental Dirac points emerging in the polariton spectrum which signifies a topological phase transition. Thus, for a fixed arrangement of the dipole emitters/antennas, we show that one can invert the sign of the valley-Chern numbers and thus switch the propagation direction of the topological polariton edge states by varying only the cavity width.

%------------FIGURE 1-------------% 
\begin{figure*}[t]
\includegraphics[width=\textwidth]{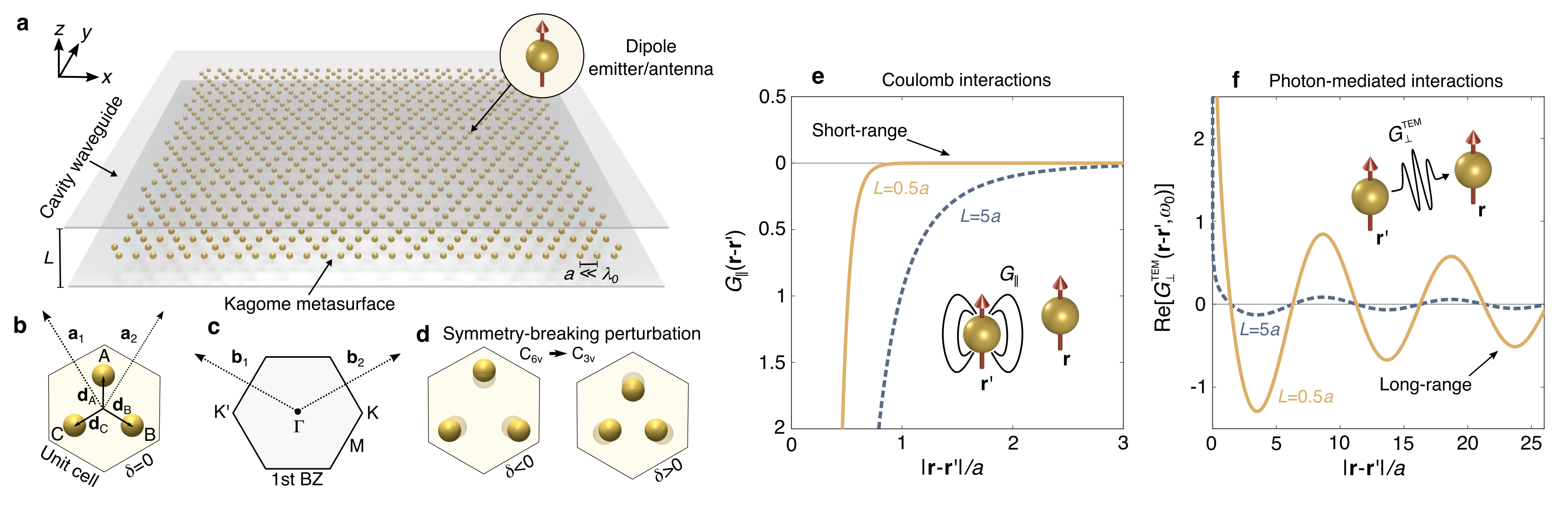}
\caption{\textbf{Symmetry-broken kagome metasurfaces.} Schematic of a kagome metasurface composed of dipole emitters/antennas with subwavelength nearest-neighbour separation $a\ll\lambda_0$, which is embedded inside a cavity waveguide of width $L$. \textbf{b} Corresponding unit cell and \textbf{c} first Brillouin zone (see Methods for the lattice parameters). \textbf{d} Expanding $\delta<0$ (left) or shrinking $\delta>0$ (right) the distances between the dipoles within the unit cell reduces the symmetry of the lattice from $C_{6v}$ to $C_{3v}$ and the subgroup of the $\text{K}/\text{K}'$ valleys from $C_{3v}$ to $C_{3}$. \textbf{e} Longitudinal Green's function for a large ($L=5a$) and small ($L=0.5a$) cavity width (dashed blue and solid orange line, respectively), which mediates short-range Coulomb interactions. \textbf{f} Real part of the transverse Green's function for a large ($L=5a$) and small ($L=0.5a$) cavity width (dashed blue and solid orange line, respectively), which describes long-range coherent interactions mediated by the cavity photons. By varying the cavity width, one can tune the dominant dipolar coupling from Coulomb to photon-mediated interactions. Plots obtained with $a=0.1\lambda_0$. }
\label{fig:KagomeMetasurface}
\end{figure*}
%------------FIGURE 1-------------% 

\textbf{Minimal kagome metasurface model.---} A schematic of a kagome metasurface embedded inside a cavity waveguide is depicted in figure\,\ref{fig:KagomeMetasurface}a. We model the resonant electric dipoles with a bare polarizability of the form $\alpha_0(\omega)=2\omega_0\mu(\omega_0^2-\omega^2-\mathrm{i}\omega\gamma_{\text{nr}})^{-1}$, where $\omega_0$ is the free-space resonant frequency, $\mu$ characterizes the strength of the polarizability, and $\gamma_{\text{nr}}$ accounts for non-radiative losses. For simplicity, we assume the polarizability to be anisotropic such that the induced dipole moments point in the $z$-direction (see inset).  We have kept the model general since it could be realized with classical dipoles ranging from microwave helical antennas \cite{Mann2018,Mann2020} to plasmonic particles \cite{Weick2013,Mann2018,Lamowski2018,Downing2019}, or quantum emitters ranging from excitonic particles \cite{Yuen-Zhou2014,Yuen-Zhou2016} to atomic dipole emitters which have been attracting considerable interest \cite{Gonzalez-Tudela2015,Bettles2016,Bettles2017,Shahmoon2017,Perczel2017,Bekenstein2020,Barredo2016,deLeseleuc2019} --- our classical analysis is valid in the single-excitation subspace \cite{Svidzinsky2010}. Moreover, this model could be generalized to include more complex antennas which also exhibit large magnetic dipole and higher-order multipole moments such as split-ring resonators \cite{Kivshar2010} and dielectric Mie resonators \cite{Peng2019}.

The unit cell, depicted in figure\,\ref{fig:KagomeMetasurface}b, contains three dipoles which form the three inequivalent hexagonal sublattices in the kagome metasurface, and the corresponding Brillouin zone is shown in figure\,\ref{fig:KagomeMetasurface}c. Moreover, we consider subwavelength nearest-neighbour separation $a\ll\lambda_0$, where $\lambda_0$ is the free-space resonant wavelength, such that the $\text{K}/\text{K}'$ points lie outside the light line and thus the metasurface supports subradiant polariton states that are evanescently bound to the lattice. Finally, we embed the metasurface at the centre of a cavity waveguide of width $L$, where the cavity walls are assumed to be perfectly reflecting mirrors \cite{Mann2018,Mann2020}. The presence of the cavity renormalizes the resonant frequency of the dipoles to $\omega_{\text{cav}}$ due to the interaction with their own image dipoles (see Methods).

\textbf{Symmetry-breaking perturbation.---} The unperturbed kagome metasurface has $C_{6v}$ symmetry while the subgroup of the $\text{K}/\text{K}'$ points is $C_{3v}$; thus, the unperturbed kagome metasurface exhibits deterministic Dirac points at the $\text{K}/\text{K}'$ points \cite{Lu2014}. To lift this degeneracy and open a non-trivial band gap, we introduce a symmetry-breaking perturbation parametrized by $\delta$ which describes the fractional change in the separation distance between the dipoles within the unit cell as schematically depicted in figure\,\ref{fig:KagomeMetasurface}d. Depending on the sign of $\delta$, we have two distinct lattices: a shrunken metasurface characterized by $\delta<0$ and an expanded metasurface characterized by $\delta>0$. We note that a similar strategy has been considered in other artificial kagome lattices \cite{Ni2017,Zhang2018,Xue2019,Ni2019,Lera2019,Wong2020,Mengyao2020}. This perturbation breaks the inversion symmetry and the $\Gamma-\text{K}(\text{K}')$ mirror symmetry, which reduces the symmetry of the metasurface to $C_{3v}$ and the subgroup of the $\text{K}/\text{K}'$ valleys to $C_{3}$, thus removing the deterministic Dirac points \cite{Lu2014}.

%------------FIGURE 2-------------% 
\begin{figure*}[t]
\includegraphics[width=1\textwidth]{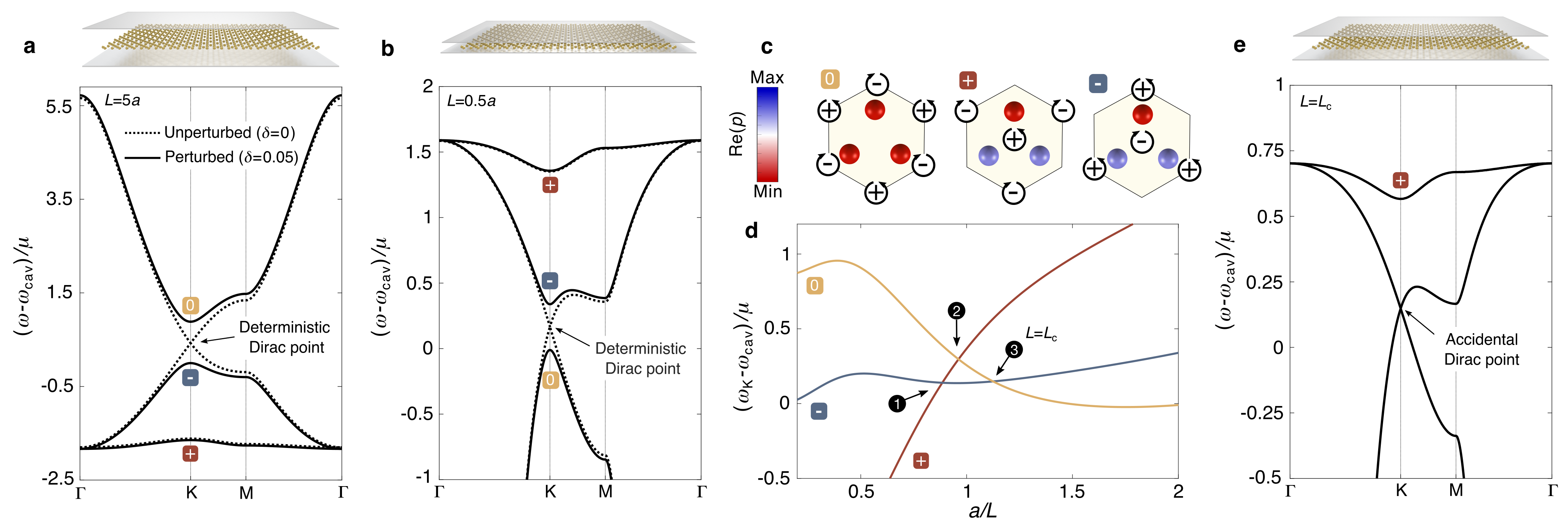}
\caption{\textbf{Cavity-induced topological band inversions.} \textbf{a}-\textbf{b} Polariton dispersion for the unperturbed kagome metasurface $\delta=0$ (dashed lines) and shrunken kagome metasurface $\delta=0.05$ (solid lines) with a large ($L=5a$) and small ($L=0.5a$) cavity width, respectively. In both regimes, the symmetry-breaking perturbation removes the deterministic Dirac points, opening a gap at the $\text{K}/\text{K}'$ points. The bands at the $\text{K}$ point are labelled according to the pseudo-angular momentum number of the eigenstates and we have schematically depicted the corresponding dipole ($p$) distributions in \textbf{c}, where $+/-$ denote the anticlockwise/clockwise phase vortices. \textbf{d} Frequency evolution of the eigenstates at the $\text{K}$ point as the cavity width is decreased for the shrunken metasurface $\delta=0.05$, where one observes three band inversions (labelled 1-3). $\textbf{e}$ Polariton dispersion at the critical cavity width $L=L_{\text{c}}$ (corresponding to the third band inversion) which exhibits accidental Dirac points. Plots obtained with $a=0.1\lambda_0$ and $\mu=0.001\omega_0$. }
\label{fig:polaritondispersion}
\end{figure*}
%------------FIGURE 2-------------% 

\textbf{Coulomb vs photon-mediated interactions.---} Crucially, the topological phase of the polaritons is not solely dictated by the intrinsic properties of the emitters/antennas and their geometrical arrangement; the topological phase also depends sensitively on the surrounding electromagnetic environment which mediates the interactions between the dipoles. These dipole-dipole interactions are encoded in the cavity Green's function $G(\mathbf{r}-\mathbf{r}',\omega)$ which describes the field generated at $\mathbf{r}$ due to a point dipole source at $\mathbf{r}'$. The Green's function can be decomposed as  $G(\mathbf{r}-\mathbf{r}',\omega)=G_{\parallel}(\mathbf{r}-\mathbf{r}')+G_{\perp}(\mathbf{r}-\mathbf{r}',\omega)$, where the longitudinal component describes the instantaneous Coulomb field and reads \cite{Mann2020}
%---------------------
\begin{equation}
    G^{}_{\parallel}(\mathbf{r}-\mathbf{r}')=-\frac{4a^3}{L}\sum_{m=1}^{\infty}k_{m}^2\operatorname{K}_0(k_{m}|\mathbf{r}-\mathbf{r}'|)\,,
    \label{eq:CoulombGreensFunction}
\end{equation}
%---------------------
where $k_{m}=2 m\pi/L$ and $\operatorname{K}_0^{}$ is the modified Bessel function of zeroth order and second kind which decays like $\operatorname{K}_0(x)\sim\mathrm{e}^{-x}/ \sqrt{x}$. The longitudinal component thus gives rise to short-range Coulomb interactions between the dipoles whose strength decreases rapidly with the separation distance as shown in figure\,\ref{fig:KagomeMetasurface}e. Furthermore, the dipoles also couple to the dynamic, transverse photonic modes of the cavity waveguide. Since we are interested in the regime of cavity widths $L<\lambda_0$, we retain only the dominant contribution from the fundamental TEM mode which gives rise to far-field radiation. The corresponding transverse Green's function reads \cite{Mann2020}
%---------------------------------------------
\begin{equation}
\begin{split}
    G^{\text{TEM}}_{\perp}(\mathbf{r}-\mathbf{r}',\omega)=\frac{\mathrm{i}\pi a^3 k_{\omega}^2}{ L}\operatorname{H}_0^{(1)}\left(k_{\omega}|\mathbf{r}-\mathbf{r}'|\right)\,,
    \end{split}
    \label{eq:TEMGreensFunction}
\end{equation}
%---------------------------------------------
where $k_{\omega}=\omega/c$ and $\operatorname{H}_0^{(1)}$ is the Hankel function of zeroth order and first kind which decays like $\operatorname{H}_0^{(1)}(x)\sim\mathrm{e}^{\mathrm{i}x}/ \sqrt{x}$. Therefore the real part of the transverse Green's function describes coherent long-range interactions mediated by the cavity photons whose strength oscillates and decreases slowly with the separation distance as shown in figure\,\ref{fig:KagomeMetasurface}f.

For large cavity widths ($L\sim \lambda_0$), the essential physics and topology near the Dirac points is dominated by the Coulomb interactions given the subwavelength spacing of metasurface (see dashed blue lines in figures\,\ref{fig:KagomeMetasurface}e-f). However, for small cavity widths ($L\ll\lambda_0$), the Coulomb interactions are exponentially suppressed and the photon-mediated interactions become dominant (see solid orange lines in figures\,\ref{fig:KagomeMetasurface}e-f).  In what follows we will see that this transition of the dominant dipolar coupling from short-range Coulomb interactions to long-range photon-mediated interactions drastically modifies the topology of the polaritons.

\textbf{Cavity-induced topological band inversions.---}  In figure\,\ref{fig:polaritondispersion}a we show the polariton dispersion (solid line) for the shrunken metasurface ($\delta=0.05$) with a large cavity width ($L=5a$) where the Coulomb interactions are dominant. Note, here we have neglected the weak photon-mediated interactions which have a negligible effect near the $\text{K}/\text{K}'$ valleys (see Supplementary Section 1). In figure\,\ref{fig:polaritondispersion}b we show the polariton dispersion (solid line) for the \emph{same} shrunken metasurface but now with a small cavity width ($L=0.5a$) where the photon-mediated interactions are dominant. In both interaction regimes, the symmetry-breaking perturbation removes the deterministic Dirac points exhibited by the unperturbed metasurface (dashed lines in figures\,\ref{fig:polaritondispersion}a-b), thus opening a gap at the  $\text{K}/\text{K}'$ points.

%------------FIGURE 3-------------% 
\begin{figure*}[t]
\includegraphics[width=0.8\textwidth]{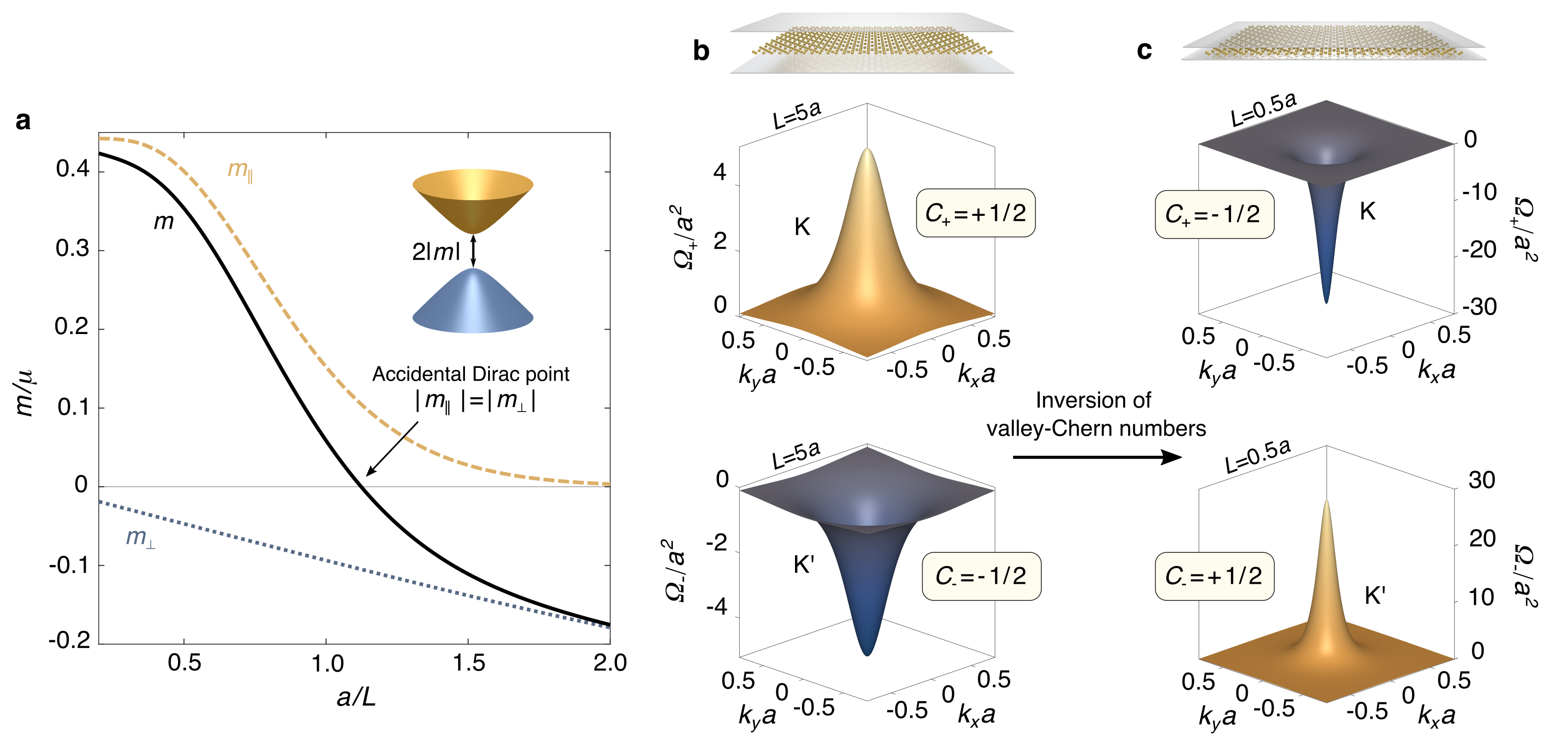}
\caption{\textbf{Cavity-induced inversion of the valley-Chern numbers.} \textbf{a} Evolution of the longitudinal (dashed orange line), transverse (dotted blue line), and total (solid black line) Dirac masses as the cavity width is reduced for the shrunken metasurface $\delta=0.05$. At the critical cavity width $L=L_{\text{c}}$, the longitudinal and transverse masses cancel each other giving rise to accidental Dirac points, which results in a topological transition where the Dirac mass switches sign. \textbf{b} Localized Berry curvature near the $\text{K}$ (top) and $\text{K}'$ (bottom) valleys and the corresponding valley-Chern numbers for a large cavity width ($L=5a$). \textbf{c} Same as \textbf{b} but with a small cavity width ($L=0.5a$). The transition from Coulomb to photon-mediated interactions switches the sign of the Berry curvature and thus inverts the valley-Chern numbers. Plots obtained with $a=0.1\lambda_0$ and $\mu=0.001\omega_0$.}
\label{fig:BandInversion}
\end{figure*}
%------------FIGURE 3-------------% 

Since the perturbation preserves the $C_3$ symmetry of the lattice, the eigenstates at the $\text{K}/\text{K}'$ points are simultaneous eigenstates of the $C_3$ operator with eigenvalues $\mathrm{e}^{\mathrm{i}\frac{2\pi}{3}l}$, where $l=0,\pm1$ are the corresponding pseudo-angular momentum numbers (see Methods). In figure\,\ref{fig:polaritondispersion}c we schematically depict the dipole distributions corresponding to the eigenstates at the $\text{K}$ point, where we have also depicted the anticlockwise/clockwise phase vortices with $+/-$.

In figures\,\ref{fig:polaritondispersion}a-b we have labelled the polariton bands at the $\text{K}$ point according to the pseudo-angular momentum number of the eigenstates. Interestingly, for the \emph{same} symmetry-breaking perturbation (i.e., a given value of $\delta$), the ordering of the eigenstates is entirely reversed for the two limiting interaction regimes. To elucidate this further, in figure\,\ref{fig:polaritondispersion}d we show the frequency evolution of the eigenstates at the $\text{K}$ point as the cavity width is progressively decreased. Due to the competition between the Coulomb and photon-mediated interactions we observe three band inversions. The first two band inversions result in $l=+1$ eigenstate evolving from being the lowest in frequency to the highest. However, here we are particularly interested in the third band inversion. At this critical cavity width $L=L_c$, the degeneracy between the $l=0$ and $l=-1$ eigenstates is restored despite the broken lattice symmetry of the perturbed metasurface, and therefore the polariton dispersion exhibits accidental Dirac points as shown in figure\,\ref{fig:polaritondispersion}e.

\textbf{Cavity-induced inversion of valley-Chern numbers.---} To characterize the topology of the third band inversion, we have derived an effective two-band Hamiltonian describing the polaritons near the $\text{K}$ ($\tau=+1$) and $\text{K}'$ ($\tau=-1$) valley in the subspace spanned by the degenerate eigenstates of the unperturbed metasurface, where $\tau=\pm1$ is the valley index (see Methods for the derivation). The effective Hamiltonian reads ($\hbar=1$)
%---------------------------------------------
\begin{equation}
    \mathcal{H}_{\text{eff}}^{\tau}=\omega_{\text{D}}(L) \mathbb{1}+v_{\text{D}}(L)(\tau\sigma_x k_x+\sigma_y k_y)+m(L)\sigma_z\,,
    \label{eq:EffectiveHamiltonian}
\end{equation}
%---------------------------------------------
where $\mathbb{1}$ is the $2\times2$ identity matrix, $\sigma_i$ are the Pauli matrices, and $\mathbf{k}=(k_x,k_y)$ is the wavevector measured from the $\text{K}/\text{K}'$ point. Note, the effective Hamiltonian\,\eqref{eq:EffectiveHamiltonian} is only valid when the other band is well separated in frequency and thus does not accurately describe the polaritons during the first two band inversions. Equation\,\eqref{eq:EffectiveHamiltonian} is equivalent to a massive Dirac Hamiltonian with a gapped spectrum $\omega=\omega_{\text{D}}\pm\sqrt{v_{\text{D}}^2k^2+m^2}$, where $\omega_{\text{D}}$ is the Dirac frequency, $v_{\text{D}}$ is the Dirac velocity, and $m$ is the Dirac mass (see Methods for the analytical expressions). 

We split the Dirac mass into two contributions $m=m_\parallel+m_\perp$, where the longitudinal mass $m_\parallel$ encodes the gap induced by the Coulomb interactions, and the transverse mass $m_\perp$ encodes the gap induced by the photon-mediated interactions. For the unperturbed metasurface ($\delta=0$), the $C_{3v}$ symmetry of the $\text{K}/\text{K}'$ valleys enforces that both the longitudinal and transverse masses separately vanish $m_{\parallel}=m_{\perp}=0$, resulting in deterministic Dirac points ($m=0$) for \emph{all} cavity widths. In stark contrast, for the symmetry-broken metasurface ($\delta\neq0$) the longitudinal and transverse masses do not vanish for \emph{any} cavity width. In figure\,\ref{fig:BandInversion}a we show how the longitudinal (dashed orange line), transverse (dotted blue line), and total (solid black line) masses vary with the cavity width for a fixed symmetry-breaking parameter ($\delta=0.05$). Crucially, the longitudinal and transverse masses have opposite signs and thus tend to compensate each other. Moreover, the longitudinal mass dominates for large cavity widths, while the transverse mass dominates for small cavity widths. At the critical cavity width $L=L_{\text{c}}$, the magnitudes of the longitudinal and transverse masses are identical ($|m_{\parallel}|=|m_{\perp}|$) and thus perfectly cancel each other ($m=0$), resulting in accidental Dirac points as shown in figure\,\ref{fig:polaritondispersion}e. Therefore, by varying only the cavity width, one can induce a topological transition where the band gap closes and reopens, switching the sign of the Dirac mass.

%------------FIGURE 4-------------% 
\begin{figure*}[t]
\includegraphics[width=1\textwidth]{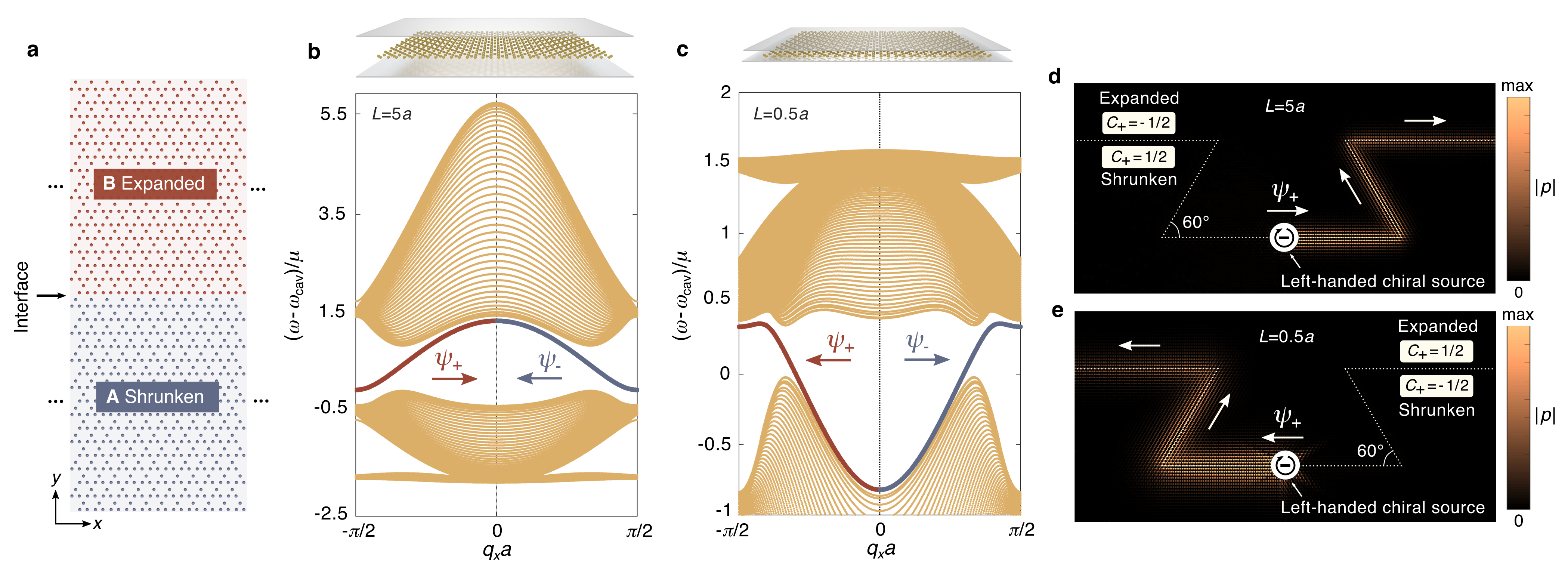}
\caption{\textbf{Switching the chirality of the topological edge states.} \textbf{a} Schematic of an interface between a shrunken metasurface $\delta_{\mathrm{A}}=0.05$ (region A) and an expanded metasurface $\delta_{\mathrm{B}}=-0.05$ (region B). \textbf{b}-\textbf{c} Interface dispersion with a large ($L=5a$) and small ($L=0.5a$) cavity width, respectively. The orange bands represent the projected bulk bands, while the red/blue bands represent the dispersion of the chiral edge states $\psi_{+}/\psi_{-}$ in the $\text{K}/\text{K}'$ valleys which propagate in opposite directions. The transition from Coulomb interactions to photon-mediated interactions switches the propagation direction of the topological edge states. \textbf{d}-\textbf{e} Simulation of the topological edge states at an interface with sharp $60\degree$  bends (dotted white lines) for a large ($L=5a$) and small ($L=0.5a$) cavity width, respectively. We excite the metasurface with a left-handed chiral source near the interface such that it selectively couples to the $\text{K}$ valley edge state, and the driving frequency is chosen to correspond with the middle of the bulk band gap. The topological edge states propagate in opposite directions due to the cavity-induced inversion of the valley-Chern numbers, thus verifying the ability to switch the chirality of the topological edge states by varying only the cavity width. Plots obtained with $a=0.1\lambda_0$ and $\mu=0.001\omega_0$. }
\label{fig:EdgeStates}
\end{figure*}
%------------FIGURE 4-------------% 

Using the eigenstates of the effective Hamiltonian\,\eqref{eq:EffectiveHamiltonian}, we calculate the localized Berry curvature (see Methods for the derivation)
%---------------------
\begin{equation}
\begin{split}
    \varOmega_{\tau}(\mathbf{k})=\tau\operatorname{sgn}(m)\frac{|m||v_{\text{D}}|^2}{2(v_{\text{D}}^2k^2+m^2)^{3/2}}
  \end{split}
  \label{eq:BerryCurvature}
\end{equation}
%---------------------
which captures the essential topology near the $\text{K}/\text{K}'$ valleys when the other band is well separated (see Supplementary Section 2). The VHTI phase of the symmetry-broken metasurface can then be characterized by non-trivial valley-Chern numbers, which for the lower band are given by
%---------------------
\begin{equation}
\begin{split}
   C_{\tau}=\frac{1}{2\pi}\int_{}\,\varOmega_{\tau}(\mathbf{k})\,\mathrm{d}^2\mathbf{k}=\frac{\tau\operatorname{sgn}(m)}{2}.
  \end{split}
  \label{}
\end{equation}
%---------------------
In figure\,\ref{fig:BandInversion}b we show the Berry curvature near the $\text{K}/\text{K}'$ valleys and their respective valley-Chern numbers for the shrunken metasurface ($\delta=0.05$) with a large cavity width ($L=5a$). The valley-Chern numbers have opposite signs for the $\text{K}/\text{K}'$ valleys and therefore the total Chern number vanishes ($C_+ +C_-=0$) as required by time-reversal symmetry. In figure\,\ref{fig:BandInversion}c we show the Berry curvature and the corresponding valley-Chern numbers for the \emph{same} shrunken metasurface but now with a small cavity width ($L=0.5a$). Remarkably, the Berry curvature within each valley changes sign due to the change in sign of the Dirac mass and, as a result, the valley-Chern numbers become inverted.

\textbf{Switching the chirality of the topological edge states.---} To see how this topological transition affects the corresponding topological polariton edge states, we consider the interface shown in figure\,\ref{fig:EdgeStates}a where region A is a shrunken metasurface ($\delta_\mathrm{A}=0.05$) and region B is an expanded metasurface ($\delta_\mathrm{B}=-0.05$). While the bulk polariton dispersion for the two regions are identical, the Dirac masses have opposite signs $m_{\mathrm{B}}=-m_{\mathrm{A}}$ and thus the interface separates regions with opposite valley-Chern numbers. As a result, each valley supports a chiral edge state that is localized at the interface with a linear spectrum $\omega_{\tau}=\omega_{\text{D}}+\operatorname{sgn}(\Delta C_{\tau})|v_{\text{D}}|k_x$ and group velocity 
%---------------------
\begin{equation}
\begin{split}
v_{\tau}=\frac{\partial \omega_{\tau}}{\partial k_x}=\operatorname{sgn}(\Delta C_{\tau}) |v_{\text{D}}|\,,
\end{split}
  \label{eq:groupvelocity}
\end{equation}
%---------------------
where $\Delta C_{\tau} = C_{\tau}^{\mathrm{A}}-C_{\tau}^{\mathrm{B}}$ is the change in valley-Chern number across the interface (see Methods). Since $\Delta C_{\tau}$ has opposite signs for the two valleys, the corresponding polariton edge states must propagate in opposite directions as required by time-reversal symmetry. Crucially, the ability to invert the sign of the valley-Chern numbers in both regions by modifying the cavity width enables one to switch the propagation direction of the topological polariton edge-states.

To verify this analytical prediction, in figure\,\ref{fig:EdgeStates}b we show the dispersion for the interface with a large cavity width ($L=5a$), where we have neglected the weak photon-mediated interactions for clarity (see Supplementary Section 1). The orange bands correspond to the projected bulk states, while the red/blue bands correspond to the spectrum of the edge states in the $\text{K}/\text{K}'$ valley. One observes that the $\text{K}$ valley edge state ($\psi_{+}$) propagates to the right while the $\text{K}'$ valley edge state ($\psi_{-}$) propagates to the left, as required by time-reversal symmetry. In figure\,\ref{fig:EdgeStates}c we show the dispersion for the \emph{same} interface but now with a small cavity width ($L=0.5a$). Despite the interface geometry being fixed, we indeed observe that the propagation direction of the topological edge states is reversed as predicted by equation\,\eqref{eq:groupvelocity}. 

To further verify this chirality inversion and demonstrate their topological robustness, we simulate the propagation of the edge states at an interface with multiple sharp $60\degree$ bends. We excite the metasurface with a left-handed chiral source and position it close to the interface such that it selectively couples to states in the $\text{K}$ valley (see Methods). In figure\,\ref{fig:EdgeStates}d, we plot the steady-state amplitude of the dipole moments for a large cavity width ($L=5a$). One clearly observes the chiral nature of the topological edge state which only propagates to the right and is not backscattered by the sharp bends in the interface. Figure\,\ref{fig:EdgeStates}e shows the simulation for the \emph{same} interface but now with a small cavity width ($L=0.5a$). As predicted, the topological edge state now propagates in the opposite direction due to the cavity-induced inversion of the valley-Chern numbers, thus verifying the ability to switch the chirality of the topological edge states by varying only the cavity width.

\textbf{Outlook.---} We have shown that the topological phase of a metasurface is not inherently fixed by design, but it can depend sensitively on the electromagnetic environment in which the metasurface is embedded. While we have demonstrated that one can induce topological transitions using a cavity waveguide, we stress that there are many alternative ways to modify the nature of the interactions between emitters/antennas by engineering different electromagnetic environments; for example, one could consider interactions mediated by surface plasmons \cite{Nikitin2011} or guided modes of photonic crystals \cite{Gonzalez-Tudela2015,Perczel2020} which may provide alternative mechanisms to induce topological transitions in arrays of interacting emitters/antennas. 

Furthermore, the underlying principle presented in this work could also have crucial implications for other topological phases that emerge in arrays of interacting emitters/antennas. For example, higher-order topological insulator phases have recently been explored in photonic kagome lattices, where it is asserted that the expanded lattice is the topological phase that exhibits topological corner states \cite{Mengyao2020}. However, here we have shown the ability to induce multiple band inversions without ever modifying the lattice geometry of the metasurface; do these band inversions switch the higher-order topological phase of the expanded/shrunken lattices? If so, it would be interesting to explore the fate of the topological corner states as the electromagnetic environment is modified.

%==================================================
%-------------------------------REFERENCES---------------------------------%
%==================================================

\begin{footnotesize}

%=============================================
\subsection{Acknowledgments}
%=============================================

\noindent C.-R.M acknowledges financial support from the Rank Prize Funds and the Engineering and Physical Sciences Research Council (EPSRC) of the United Kingdom through the EPSRC Centre for Doctoral Training in Metamaterials (Grant No. EP/L015331/1). E.M. acknowledges financial support from the Royal Society International Exchanges Grant IEC/R2/192166.

%=============================================
\subsection{Author contributions}
%=============================================

\noindent C.-R.M. conceived the idea, developed the theory, performed the calculations, and wrote the manuscript; E.M. contributed to the theoretical understanding, and supervised the project; All authors commented on the manuscript.

%=============================================
\subsection{Data availability}
%=============================================

\noindent All relevant data are available from the corresponding authors upon reasonable request.

%=============================================
\subsection{Competing interests}
%=============================================

\noindent The authors declare no competing interests.

%=============================================
\subsection{Additional information}
%=============================================

\noindent Correspondence and requests for materials should be addressed to C.-R.M.

\clearpage

\end{footnotesize}

\end{document}